\documentclass[aps,prd,reprint,groupedaddress]{revtex4-2} 
\usepackage{graphicx}
\usepackage{verbatim}
\usepackage{url}
\usepackage{tabularx}
\usepackage{amsmath}
\usepackage{xcolor} 
\usepackage{hyperref}  
\hypersetup{
colorlinks=false}
\graphicspath{{image/}{image/bmu-compare}{image/cooling-curve}{image/relationship-image}}

\begin{document}

\title{Thermal emission from dark matter–heated neutron stars in the Galactic Center}

\author{Yunhe Hu}
\affiliation{Department of Astronomy, Xiamen University, Xiamen, Fujian 361005, China}
\author{Feng Huang}
\email{fenghuang@xmu.edu.cn}
\affiliation{Department of Astronomy, Xiamen University, Xiamen, Fujian 361005, China}
\author{Taotao Fang}
\affiliation{Department of Astronomy, Xiamen University, Xiamen, Fujian 361005, China}

\date{\today}

\begin{abstract}

We investigate the thermal impact of dark matter (DM) capture and annihilation on neutron stars (NSs) in the Galactic Center (GC). Accounting for both kinetic energy deposition and internal annihilation, we systematically evaluate the influence of various DM density profiles, ranging from cored to cuspy distributions, on the late-time thermal evolution of NSs. For NSs older than $\sim 10^7~\mathrm{yr}$, the surface temperature approaches an equilibrium value $T_\mathrm{s}^{\mathrm{eq}} \sim 10^4$--$10^6~\mathrm{K}$, depending on the stellar location and the ambient DM density. In the presence of a density spike, enhanced heating shifts the emission toward ultraviolet (UV) and soft x-ray bands; however, strong interstellar extinction and large hydrogen column densities significantly suppress the observable flux density. We further provide an estimate of the cumulative infrared surface brightness from the NS population in the GC. The predicted flux density from an individual NS remains below $\sim 0.1\,\mathrm{nJy}$, while the integrated emission yields an average surface brightness $I_\nu \lesssim 10^{-9}\,\mathrm{Jy\,arcsec^{-2}}$, corresponding to a signal-to-noise ratio well below current detection thresholds. 
Our results indicate that thermal signatures from DM-heated NSs in the GC remain below the sensitivity limits of current instruments, although nearby systems with lower extinction may provide more promising targets for detection
\end{abstract}

\maketitle

\section{Introduction} \label{sec:intro}


Cosmological models and observations suggest that galaxies are embedded within dark matter (DM) halos, with densities that increase toward their centers. As the closest such region, the Galactic Center (GC) has long been regarded as a promising target for DM searches. Over the past decade, the GeV excess detected within a few degrees of the GC by the Fermi Large Area Telescope (Fermi-LAT) has attracted considerable attention as a potential signature of DM annihilation~\cite{2016_Linden_CharcterizationGCEXCESS, 2009_GOODEN_Hooper_GCEXCESS}. However, the relative distributions of DM and baryonic matter in the GC remain debated; while DM dominates the overall mass of the Galaxy, baryonic matter is thought to prevail in the innermost regions~\cite{2014_Timlinden_baryonGC}. A population of unresolved millisecond pulsars, whose gamma-ray emission is expected to originate primarily from curvature radiation, has been widely considered as a competitive interpretation of this excess~\cite{2011_Abazajian_consistency}. Despite extensive study, neither the DM nor the pulsar origin has been conclusively confirmed~\cite{2013_Mirabal_DMversuspulsar, 2014_BramanteJoseph_DMGCPulsar}.

Given the high concentration of both DM and baryonic matter in the GC, the interaction between DM particles and resident celestial bodies has attracted significant interest. In this dense environment, DM particles frequently traverse stellar interiors, where elastic scattering with stellar constituents transfers kinetic energy and renders the particles gravitationally bound to the host star~\cite{1989_Itzhak_WIMPmass_prd}. This continuous capture leads to the gradual accumulation and subsequent annihilation of DM within the stellar core. The resulting energy deposition can alter the object’s thermal balance and evolutionary track, potentially manifesting as observable anomalies in luminosity or cooling behavior. Quantifying these signatures provides a powerful means to probe DM-baryon interactions under extreme astrophysical conditions.

During the last decade, driven by the increasing number and improved quality of GC observations, research on DM annihilation in celestial bodies has expanded across various astrophysical scales~\cite{2024_Bramante_Raj_reviewDMcpstar,2022_Baryakhtar_caputo_reviewExtre}. Orbital dynamics studies suggest that DM accretion can alter the decay rate of pulsar binaries~\cite{2025_Mishra_OrbitalPulsar} or even trigger the collapse of NSs into black holes in the GC~\cite{2023_Dicongliangshaolijing_improvedbounds}. In terms of stellar evolution, DM annihilation heating in gas clouds---such as the G2 cloud---has been proposed as a novel way to constrain the DM parameter space~\cite{2023_Manhochan_ANewmethod}. DM annihilation may also disrupt stars or prolong their lifetimes, leading to a ``dark main sequence” on the Hertzsprung-Russell diagram~\cite{2024_John_DMscatt,2025_John_Rebecca_Immortal}. Similarly, brown dwarfs powered by DM annihilation heating may exhibit reduced lithium depletion; such “dark dwarfs"~\cite{2025_DjunaCroon_Darkdwarfs} are expected to reside near the GC and remain to be discovered. Furthermore, heating from DM can limit hydrogen and helium accretion onto protoplanetary cores, potentially preventing the formation of Jovian planets and implying that exoplanets may serve as natural detectors of DM~\cite{2021_Leane_Smirnov_exoplanets, 2024_Croon_youngplanet}. Together, these phenomena demonstrate the broad impact of DM on the thermal and dynamical states of stellar objects.

Among these celestial probes, neutron stars (NSs) stand out as particularly sensitive detectors due to their extreme baryonic densities and deep gravitational potentials. These properties allow NSs to efficiently capture DM particles even at suppressed scattering cross sections~\cite{2021_LeaneRebeccaK_Celestialthrg}. As infalling DM is accelerated to relativistic speeds, its kinetic energy is transferred to the stellar interior through collisions, providing a significant heating component~\cite{2017_Baryakhtar_kwindows}. Combined with the heat from subsequent DM annihilation, this process can alter the standard cooling of the star, maintaining a nearly constant surface temperature over long timescales~\cite{2008_Kouvaris_col, 2008_Bertone_CSDMprobe, 2010_Arnaud_Fairbairn_NSprob}. Consequently, the thermal evolution of NSs in the GC could potentially to be a sensitive indicator of the local DM environment.

The GC hosts a dense and complex stellar environment, widely suggested to contain a substantial population of NSs. Although their exact distribution and abundance remain debated, mounting observational and theoretical evidence supports the existence of such a population~\cite{2003_Genzel_OBstar,2003_Levin_Belobor_OBremant}. Theoretically, dynamical friction and mass segregation are expected to preferentially concentrate compact objects toward the central regions~\cite{2006_Freitag_Amaro_seg,2006_Hopman_Alexander_segra1apj,2016_Aharon_Perets_segra2prd}.
Observational searches for supernova remnants, pulsars, and low-mass x-ray binaries have been reviewed in Ref.~\cite{2018_Kim_Davies_NSreview_JKAS}. While previous research has explored cumulative signals from these populations—particularly high-energy neutrino and gamma-ray emissions from DM annihilating outside stars via long-lived mediators~\cite{2021_LeaneRebeccaK_Celestialthrg, 2022_BoseDebajit_popula, 2024_Aceevedo_Santos_WDdetector}—a systematic investigation of the thermal impact from DM annihilation inside the NSs remains lacking. This gap is particularly significant because the heating rate depends sensitively both on the spatial distribution of NSs and the ambient density of DM. In the GC, the latter may vary by up to seven orders of magnitude depending on the assumed density profile, from extreme cuspy spikes to conservative cored models~\cite{2001_Ullio_Zhao_spikeprd, 2023_Shyam_silk_DMspike, 2024_Shen_Yuan_SPorbits}. Consequently, stellar location plays a decisive role in determining the resulting thermal signature. In this work, we revisit the DM-induced heating and the subsequent thermal evolution of NSs across the GC. We quantify the thermal properties and observational detectability of both individual sources and the projected NS population across multiple wavebands.

The paper is organized as follows. In Sec. II, we describe the physical framework of the capture, cooling and heating processes, alongside the adopted DM density profiles. In Sec. III, we present our numerical results for NS cooling curves, blackbody luminosities, and flux densities. Finally, a summary of our main conclusions is provided in Sec. IV.

\section{Physical Framework}
\label{sec:floats}
When a NS moves through a DM halo, DM particles may scatter off the stellar constituents. 
If a DM particle loses enough energy to drop below the local escape velocity $v_{\mathrm{esc}}$, it becomes gravitationally captured. DM particles can subsequently annihilate and deposit energy, affecting the thermal evolution of the star. 
\subsection{Capture and annihilation}
As DM particles traverse the compact star, such as a NS, each particle is assumed to undergo at most one scatter with a nucleon, provided its mean free path is larger than the stellar radius. We consider the regime in which a single scattering is sufficient for capture. The capture rate per unit volume in the rest frame of the NS can be written as~\cite{1987_Gouldandrew_Resonantwinp, 2012_Mcdermottsamuel_asyDM},
\begin{equation}
\begin{aligned}
    \frac{dC(r)}{dV}=&\,\sqrt{\frac{6}{\pi}}\,n_{\chi}(r)\,n_{\mathrm{B}}(r)\, \xi\, 
    \frac{v_{\mathrm{esc}}(r)^2}{\bar{v}^2}
\left(\bar{v}\sigma_{\chi\mathrm{_B}}\right)\\
    &\times\left[1-\frac{1-\exp(-A^2)}{A^2}\right].
\end{aligned}
\label{eq:capture_with_volume}
\end{equation}
Here, $n_{\chi}(r)$ and $n_{\mathrm{B}}(r)$ are the number densities of DM particles and stellar baryons, respectively; $v_{\mathrm{esc}}(r)$ is the DM escape velocity at the radius $r$; and $\bar{v}$ is the DM velocity dispersion in the ambient compact star environment. $\sigma_{\chi\mathrm{_B}}$ is the effective scattering cross section between DM particles and nucleons, and it is constrained by direct-detection experiments \cite{2016_Marrod_diriew}. 
 The factor $\xi$ accounts for the suppression due to nucleon degeneracy. 
If DM mass $m_\chi$ is larger than the nucleon mass ($m_\chi \gtrsim \mathrm{1 GeV}$), all nucleons can participate in the scattering process, and one safely takes $\xi = 1$~\cite{2012_Mcdermottsamuel_asyDM}.
The parameter $A^2$ controls the probability of capture in a single scattering and is defined as,
\begin{equation}
     A^2 = \frac{3}{2}\frac{v_\mathrm{esc}(r)^2}{\bar{v}^2}\frac{\mu}{\mu_{-}^2},
\end{equation}
where $\mu=m_{\chi}/m_\mathrm{B}$ and $\mu_{-}=(\mu-1)/2$. Within the mass range $1~\mathrm{GeV}<m_\chi<100~\mathrm{TeV}$, $A^2 \gg1$, and the square-bracket term is effectively unity. Under these conditions, the capture probability per scattering is insensitive to $m_\chi$, and we adopt $m_\chi=100~\mathrm{GeV}$ as a representative value. Integrating over the stellar volume, the total capture rate for a NS is,
\begin{equation}
    C_{\mathrm{tot}} = 4\pi\int_{0}^{R_\mathrm{NS}}r^2\,\frac{dC(r)}{dV}\,dr. 
\end{equation}
 $R_{\mathrm{NS}}$ is the radius of NS. For simplicity, assuming constant densities $n_{\chi}$, $n_B$ and $v_\mathrm{esc}(r)=v_\mathrm{esc}$, we obtain,
\begin{equation}
C_{\mathrm{tot}}\simeq \sqrt{\frac{6}{\pi}}\frac{\rho_\chi}{m_{\chi}}\frac{v_{\mathrm{esc}}^2}{\bar{v}^2}\,(\bar{v}\sigma_{\chi\mathrm{_B}})\,N_{\mathrm{B}}\,\left[1-\frac{1-\exp(-A^2)}{A^2}\right],
\label{eq:single_capture}
\end{equation}
$\rho_\chi$ is the local DM density and $N_{\mathrm{B}}$ represents the total baryon number in the star. 
The effective scattering cross section is bounded by the geometric cross section of the NS and is given by 
\begin{equation}
\sigma_{\chi\mathrm{_B}}=\mathrm{Min}\left(\sigma_{\chi\mathrm{_B}}, \sigma_{\mathrm{geo}}\right), \quad \sigma_{\mathrm{geo}}=\frac{\pi R_{\mathrm{NS}}^2}{N_{\mathrm{B}}}.
\end{equation}
For NSs, DM capture needs to be treated within a general relativistic framework. A DM particle just grazing the stellar surface follows a geodesic in the Schwarzschild spacetime. The corresponding impact parameter is given by~\cite{1989_Itzhak_WIMPmass_prd, 2008_Kouvaris_col}
\begin{equation}
b = \left(\frac{2GM_{\mathrm{NS}}R_{\mathrm{NS}}} {v_{\mathrm{esc}}^2}\right)^{1/2}\left[1-2GM_{\mathrm{NS}}/R_{\mathrm{NS}}\right]^{-1/2},
\end{equation}
where the term in square brackets accounts for general-relativistic corrections due to spacetime curvature. Since the capture rate scales with the effective geometric cross section $\pi b^2$, the total DM capture rate is enhanced by a relativistic focusing factor~\cite{2017_Bramante_Delgado_mulscatter},
\begin{equation}
    C_{\mathrm{tot}}
\rightarrow
\frac{1}{1-2GM_{\mathrm{NS}}/R_{\mathrm{NS}}}
\,C_{\mathrm{tot}}.
\end{equation}
In the above expressions, we adopt natural units with the speed of light $c=1$.

As DM accumulates in the stellar core, the annihilation rate rises continuously. The number of DM particles $N_\chi(t)$ satisfies,
\begin{equation}
\frac{dN_\chi(t)}{dt}=C_{\mathrm{tot}}-C_{\mathrm{ann}}N_\chi(t)^2,
\label{eq:differential_number}
\end{equation}
where the annihilation coefficient is,
\begin{equation}
C_{\mathrm{ann}}=\left \langle \sigma_{\mathrm{ann}}v \right \rangle /V  
\end{equation}
with $ \left \langle \sigma_{\mathrm{ann}}v \right \rangle$ being the velocity-averaged annihilation cross section.
 The solution of the differential equation is,
 \begin{equation}
     N_\chi(t)=\sqrt{\frac{C_{\mathrm{tot}}}{C_{\mathrm{ann}}}}\tanh\frac{t}{t_{\mathrm{eq}}}, 
 \end{equation}where 
 \begin{equation}
     t_{\mathrm{eq}}=1/\sqrt{C_{\mathrm{ann}}C_{\mathrm{tot}}}
 \end{equation}
 is the timescale to reach equilibrium between capture and annihilation. 
The annihilation rate at equilibrium is,
\begin{equation}
    \Gamma_{\mathrm{ann}}= \frac{1}{2}C _{\mathrm{ann}} N_\chi^2=\frac{1}{2}C_{\mathrm{tot}}.
    \label{eq:equilibrium_gamma}
\end{equation}
The applicability of this equilibrium condition relies on the age of NSs. NSs located in the GC are generally considered to be old, with typical ages exceeding 5 Gyr~\cite{Generozov_2018}. 
The equilibrium between capture and annihilation is reached well before DM is fully thermalized with the NS, which occurs on a short timescale of less than $10^3$ years~\cite{2024_Bell_ThermalNS}. Thus, we can safely assume that the equilibrium has been established.

\subsection{NS cooling and DM heating} 

A nascent NS is born with an extremely high temperature of approximately $10^{11}~\mathrm{K}$. As the NS evolves, it cools through neutrino emission and photon radiation. Since these cooling processes scale with temperature as a power law, their efficiency decreases rapidly as the NS temperature drops. At sufficiently low temperatures, the heating generated by DM could therefore dominate the thermal evolution of the NS.

The energy injected into the NS by DM consists of two distinct
contributions. The first is the annihilation energy,
\begin{equation}
\mathcal{E}_{\chi} = 2\Gamma_{\mathrm{ann}} m_\chi c^2,
\end{equation}
where we assume that the rest-mass energy released in DM annihilation is fully converted into heat. Generally, even if the annihilation products include neutrinos, these high-energy particles are initially trapped within the NS. Although they could eventually escape after losing most of their energy through multiple scatterings, the energy carried away is negligible and therefore excluded from our calculations.
The second contribution arises from the kinetic energy of DM particles captured by the NS. 
DM particles falling into the strong gravitational field of the NS are accelerated to relativistic speeds and subsequently lose their kinetic energy through collisions with stellar matter, given by
\begin{equation}
\mathcal{K}_{\chi} = (\gamma - 1)\,2\Gamma_{\mathrm{ann}} m_\chi c^2,
\end{equation}
where $\gamma = (1 - v_{\mathrm{esc}}^2/c^2)^{-1/2}$ is the Lorentz factor of the DM particle at the NS surface.
Accordingly, the total
DM energy emissivity is given by~\cite{2018_Chen_Lin_rehetempTaipei}
\begin{equation}
    \epsilon_\chi = \frac{\mathcal{E}_{\chi} + \mathcal{K}_{\chi}}{V}.
\end{equation}

The initial cooling of a NS is dominated by neutrino emission via the modified Urca process, which is easier than the direct Urca process, because an additional bystander nucleon absorbs the excess momentum and ensures momentum conservation. The corresponding neutrino emissivity is given by~\cite{1983_Shapiro_book, 2008_Kouvaris_col}
\begin{equation}
    \epsilon_\nu=(1.2\times 10^4\,\mathrm{erg}\,\mathrm{cm}^{-3}\mathrm{s}^{-1})\left( \frac{n_\mathrm{B}}{n_{0}} \right)^{2/3}\left(\frac{T_\mathrm{c}}{10^7\,\mathrm{K}}\right)^8,
\end{equation}
where $n_0 = 0.17~\mathrm{fm}^{-3}$ is the saturation density of nuclear matter, and $T_\mathrm{c}$ denotes the NS core temperature.

Photon emission from the NS surface provides another cooling mechanism. The photon luminosity follows the Stefan-Boltzmann law,
\begin{equation}
L_\gamma = 4\pi R_{\mathrm{NS}}^2 \sigma_{\mathrm{SB}} T_{\mathrm{s}}^4,
\label{eq:blackbody}
\end{equation}
where $\sigma_{\mathrm{SB}}$ is the Stefan-Boltzmann constant and $T_\mathrm{s}$ is the NS surface temperature. $T_{\mathrm{s}}$ is related to  $T_\mathrm{c}$ and may be approximated as \cite{1983_Gudmundsson_Epstein_Envelopes}
\begin{equation}
    T_{\mathrm{s}}=(0.87 \times10^6\,\mathrm{K}) \left(\frac{g_\mathrm{s}}{10^{14}\,\mathrm{cm/s^2}}\right)^{1/4}\left(\frac{T_\mathrm{c}}{10^8\,\mathrm{K}}\right)^{0.55}.
\label{eq:NStemrelation}
\end{equation}
Here, $g_\mathrm{s} = GM_{\mathrm{NS}}/R_{\mathrm{NS}}^2$ is the surface gravitational acceleration.
The photon emissivity for the entire NS is then
\begin{equation}
    \epsilon_{\gamma}=\frac{L_\gamma}{(4/3)\pi R_{\mathrm{NS}}^3}=1.8\times10^{14}\left(\frac{T_\mathrm{c}}{10^8\,\mathrm{K}}\right)^{2.2}\,\mathrm{erg}\,\mathrm{cm}^{-3}\,\mathrm{s}^{-1}.
\end{equation}

Combining these three processes described above, heating from DM annihilation, neutrino cooling via the modified Urca process, and photon cooling from the surface, the evolution of $T_\mathrm{c}$ is governed by,
\begin{equation}
 \frac{dT_\mathrm{c}}{dt} =\frac{-\epsilon_{\nu}-\epsilon_{\gamma}+\epsilon_{\chi}}{
c_V}.
\label{eq:dT_dt_evo}
\end{equation}
 The specific heat per unit volume, $c_V$, for degenerate and noninteracting fermion gases is,
\begin{equation}
    c_V = \frac {k_B^2T_\mathrm{c}}{3\hbar^3 c} {\sum_{i}}p^i_\mathrm{F} \sqrt{m_{i}^2c^2+(p^i_\mathrm{F})^2},
\end{equation} 
where the index $i$ runs over neutrons~(n), protons~(p), and electrons~(e). For density $n_\mathrm{B} \lesssim 2n_{0}$, the Fermi momenta are approximately 
\cite{1939_Chandrasekhar_stellarstruc}
\begin{align}
    &p^\mathrm{n}_\mathrm{F}=(340~\mathrm{MeV})\left(\frac{n_\mathrm{B}}{n_0}\right)^{1/3}, \\
    &p^\mathrm{p}_\mathrm{F}=p^\mathrm{e}_\mathrm{F}=(60~\mathrm{MeV})\left(\frac{n_\mathrm{B}}{n_0}\right)^{2/3}.
\end{align} 
Consequently, the cooling curve of a NS can be obtained by solving Eq.~\eqref{eq:dT_dt_evo}, as described in Sec.~\ref{Sec:Results}.

 \begin{figure}[htb]
     \centering
     \includegraphics[width=\linewidth]{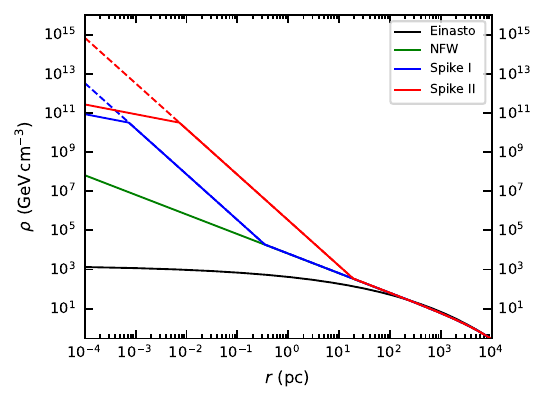}
     \caption{DM density profiles are shown as functions of the
galactic radius $r$. Shown are the Einasto (black), NFW
halo (green), and two spike profiles: Spike~I (blue, $R_\mathrm{sp}=0.34~\mathrm{pc}$) and Spike~II (red, $R_\mathrm{sp}=18.6~\mathrm{pc}$).
Solid (dashed) lines denote profiles with (without) saturation due to DM self-annihilation. All profiles are normalized to the local DM density
$\rho_\odot = 0.383~\mathrm{GeV\,cm^{-3}}$ at $R_\odot = 8.2~\mathrm{kpc}$. The saturation density assumes $m_\chi=100~\mathrm{GeV}$ and an annihilation cross section of $\left \langle \sigma v \right \rangle = 10^{-26}~\mathrm{cm^3\,s^{-1}}$.}
     \label{fig:DM_profile}
 \end{figure}
\subsection{DM density profile}
The distribution of DM in the GC is not well constrained and may exhibit either cuspy or cored behavior, depending on the effects of baryonic feedback and related astrophysical processes. We adopt the standard Navarro-Frenk-White~(NFW) halo~\cite{1996_NFW_structure} and steeper spike profiles for the cuspy case, and the Einasto profile for the cored case. For each profile, the density normalization $\rho_0$ is fixed by requiring the DM density at the solar position to match the local value $\rho_{\odot}=0.383~\mathrm{GeV\,cm^{-3}}$ at the solar radius $R_{\odot}=8.2~\mathrm{kpc}$, while the scale radius
$r_\mathrm{s}$ is model dependent.

The NFW profile was originally motivated by early collisionless cold DM $\mathrm{N}$-body simulations and is characterized by a cuspy inner density structure. The NFW halo can also be expressed in a generalized form,
  \begin{equation}
      \rho_{\mathrm{NFW}}(r)=\frac{\rho_{0}}{(r/ r_\mathrm{s})^{\gamma_\mathrm{h}}[1+(r/r_\mathrm{s})]^{3-{\gamma_{\mathrm h} }}}, 
  \end{equation}
where $r_{\mathrm s}$ is the scale radius, taken as $r_{\mathrm s} = 18.6~\mathrm{kpc}$ for the Milky Way. The inner slope $\gamma_\mathrm h$ determines the steepness of the central DM density. In this work, we adopt $\gamma_\mathrm{h} =1$, which corresponds to the standard NFW profile. In the presence of a central SMBH, the gravitational potential can significantly reshape the surrounding DM distribution, leading to the density spike in the innermost region of the NFW host halo~\cite{1999_Gondolo_Spike,2013_Sadeghian_Ferrer_DMdist_prd}. The DM spike density profile can then be written as,
\begin{equation}
   \rho_{\rm spike} = \rho_{\mathrm{NFW}}(R_{\mathrm{sp}}) \left( \frac{r}{R_{\mathrm{sp}}}\right)^{-\gamma_{\mathrm{sp}}},
\end{equation}
where $R_{\mathrm{sp}}$ denotes the characteristic spike radius, within which the DM density is significantly enhanced by the SMBH potential. The spike slope $\gamma_{\mathrm{sp}}$ is determined by 
$\gamma_\mathrm{h}$,
\begin{equation}
   \gamma_{\mathrm{sp}}
   = \frac{9 - 2\gamma_\mathrm{h}}{4 - \gamma_\mathrm{h}}=\frac{7}{3} .
\end{equation} Moreover, DM self-annihilation can eventually regulate the density, producing a saturation density and modifying the spike profile~\cite{2023_Shyam_silk_DMspike},
\begin{equation}
    \rho(r)=\left\{
\begin{array}{ll}
0                                             & r<2R_{\mathrm{sch}}  \\
\rho_{\mathrm{sat}}\left( \frac{r}{R_{\mathrm{sat}}} \right)^{-0.5}    & 2R_\mathrm{sch}\leq r<R_{\mathrm{sat}} \\
\rho_{\mathrm{spike}}(r)                               &    R_{\mathrm{sat}} \leq r<R_{\mathrm{sp}} \\  
\end{array}                                  
\right. .
\end{equation}
Here, $R_{\mathrm{sch}}$ is the Schwarzschild radius of the SMBH. Within $r\lesssim R_{\mathrm{sat}}$, DM self-annihilation becomes efficient, softening the density profile and leading to the formation of a saturation density~\cite{2023_Shyam_silk_DMspike},
\begin{equation}
    \rho_{\mathrm{sat}} = \frac{m_\chi}{\left \langle \sigma v \right \rangle t_{\mathrm{BH}}} \simeq 3.17\times10^{10}~\mathrm{GeV\,cm^{-3}}.
\end{equation}
Here, we adopt a DM mass of $m_\chi = 100~\mathrm{GeV}$ with a canonical thermal annihilation cross section $\left \langle \sigma v \right \rangle = 10^{-26}\, \mathrm{cm}^{3}\,\mathrm{s}^{-1}$ and assume a SMBH age of $t_{\mathrm {BH}}=10^{10}\,\mathrm{yr}$. $R_{\mathrm{sat}}$ is determined by solving $\rho_{\mathrm{sat}}=\rho_{\mathrm{spike}}(R_{\mathrm{sat}})$.
\begin{figure*}[htb!]
    \centering
    \includegraphics[width=\textwidth]{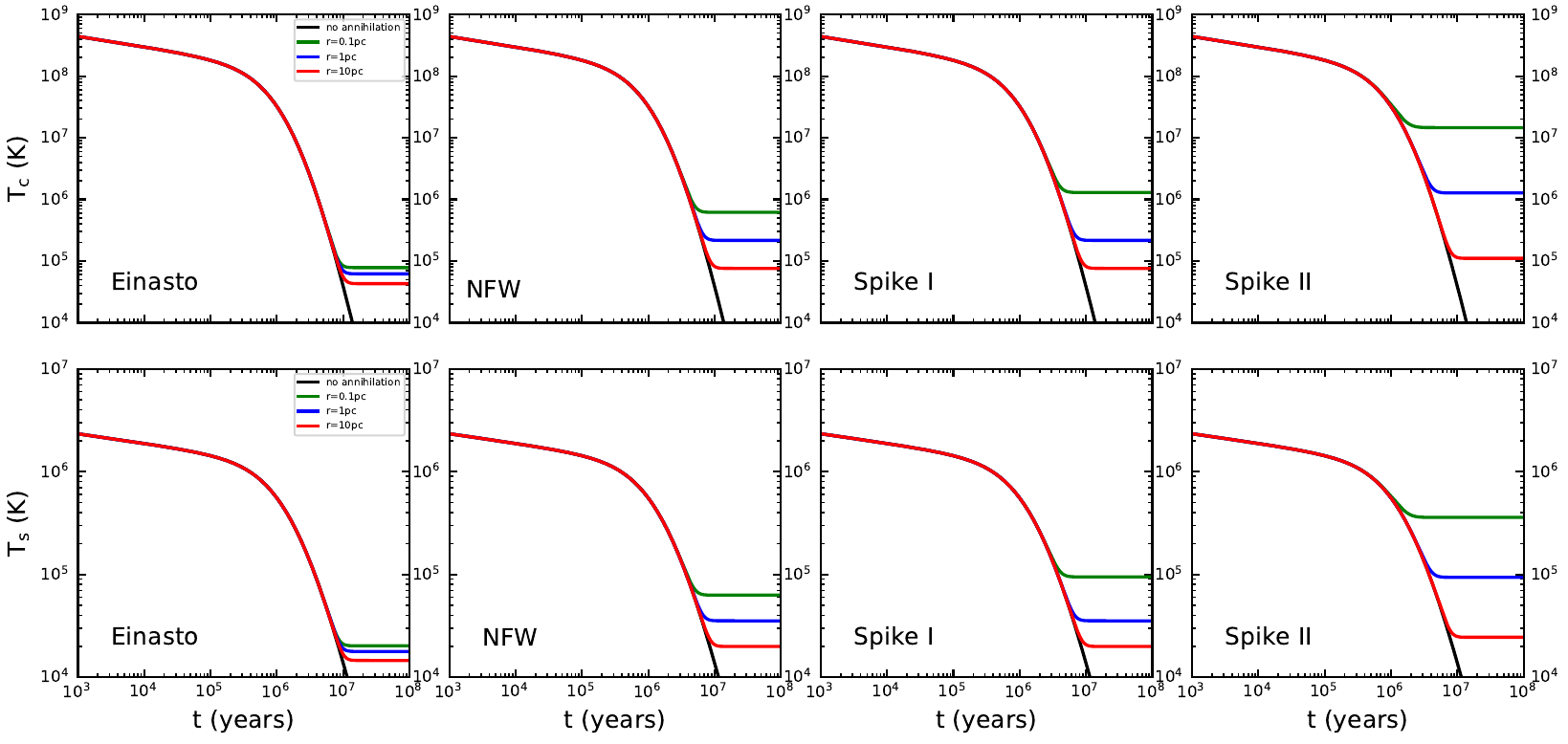}
    \caption{Cooling curves show the time evolution of the core temperature $T_{\mathrm{c}}$ (top panels) and surface temperature $T_{\mathrm{s}}$ (bottom panels) for a NS with $M_{\mathrm{NS}}=1.5~M_{\odot}$ and $R_{\mathrm{NS}}=10~\mathrm{km}$, considering four DM profiles: Einasto, NFW, Spike I, and Spike II. The colors indicate the distance from the GC: green, blue, and red correspond to $0.1~\mathrm{pc}$, $1~\mathrm{pc}$, and $10~\mathrm{pc}$. The black line represents the case neglecting DM annihilation. Calculations assume $m_\chi = 100~\mathrm{GeV}$ and $\sigma_{\chi_\mathrm{B}}=10^{-45}~\mathrm{cm^2}$.}
    \label{fig:cooling-curve}
\end{figure*}
Numerical studies~\cite{2004_Merritt_spikerh} indicate that the DM spike occurs around a radius of approximately $R_{\mathrm{sp}}\approx 0.2R_{\mathrm{h}}$, largely independent of the host DM halo profile, where $R_\mathrm{h}$ is the gravitational influence radius of the SMBH. For the Milky Way, adopting $R_{\mathrm{h}}=1.7~\mathrm{pc}$ leads to $R_{\mathrm{sp}}=0.34~\mathrm{pc}$~\cite{2023_Shyam_silk_DMspike}. However, according to Gondolo and Silk~\cite{1999_Gondolo_Spike},
 the spike radius depends on $\gamma_\mathrm{h}$ of the NFW halo and can be written as,
 \begin{equation} R_{\mathrm{sp}}=\alpha_{\gamma}r_{\mathrm s}(M_{\rm BH}/\rho_0r_{\mathrm s}^3)^{1/(3-\gamma_\mathrm{h})},
 \end{equation}
where $\alpha_{\gamma}$ is a dimensionless scale factor and $M_{\mathrm{BH}}$ is the mass of the SMBH. For $\gamma_\mathrm{h} = 1$, we adopt $R_{\mathrm{sp}}=18.6~\mathrm{pc}$, following Ref.~\cite{2024_Shen_Yuan_SPorbits}. We therefore adopt two spike scenarios, denoted as Spike I and Spike II, corresponding to the numerically motivated spike radius $R_{\mathrm{sp}}= 0.34~\mathrm{pc}$ and the analytic Gondolo and Silk spike radius $R_{\mathrm{sp}}= 18.6~\mathrm{pc}$, respectively. 

Stellar feedback processes, such as supernova explosions and stellar winds, may flatten the inner regions of the DM density profile, transitioning it from a cuspy to a core distribution~\cite{2008_Mashchenko_feedbackcore,2025_Valenciano_Coredgalaxy}. As a conservative alternative, we adopt 
the Einasto profile~\cite{1965_Einasto} to represent a cored DM distribution,
\begin{equation}
    \rho_{\mathrm{Einasto}} = \rho_{0} \exp{ \left\{-\frac{2}{\alpha}
    \left[ \left(\frac{r}{r_{\mathrm{s}}}\right)^\alpha - 1  \right]
    \right\} }
\end{equation}
with a shape parameter $\alpha=0.2$ and a scale radius $r_{\mathrm{s}} = 20~\mathrm{kpc}$ \cite{2022_BoseDebajit_popula}.

These DM density profiles are shown in Fig.~\ref{fig:DM_profile}. While the profiles are similar at large radii, significant deviations emerge within the inner $100\,\mathrm{pc}$ of the GC. Specifically, as the radius approaches $R_{\mathrm{sp}}$, the densities in the spike scenarios rise steeply. By $r=0.01~\mathrm{pc}$, the spike density exceeds the Einasto profile by a factor of $\sim 10^7$. In the innermost regions, DM self-annihilation becomes highly efficient and softens the inner density profile, leading to a saturation of the central DM density. This is in contrast to the case neglecting the effect of DM annihilation~(dashed lines), where the density continues to rise following the original power-law slope. 

\section{Results}
\label{Sec:Results}
In this Section, we present the numerical results for the thermal evolution of NSs induced by DM heating. Specifically, NSs located at different galactic radii are investigated, and the corresponding cooling curves are obtained. A benchmark model is adopted throughout this Section. We assume a NS with $M_{\mathrm{NS}}=1.5~M_{\odot}$ and $R_\mathrm{NS}=10~\mathrm{km}$, 
corresponding to a surface escape velocity of $v_{\mathrm{esc}}=1.8\times 10^{5}~\mathrm{km/s}$. The DM mass is fixed to $m_\chi = 100\, \mathrm{GeV}$, and the velocity dispersion is taken to be $\bar{v}=220~\mathrm{km/s}$. For the DM-baryon effective scattering cross-section, a representative value of $\sigma_{\mathrm{\chi_B}}=10^{-45}~\mathrm{cm}^{2}$ is adopted. This choice lies below the geometric saturation limit of the NS, $\sigma_{\mathrm{geo}} \simeq 1.8 \times 10^{-45}~\mathrm{cm}^{2}$, for the typical NS parameters adopted above, and falls within the range probed by current direct detection experiments for weakly interacting massive particles (WIMPs)~\cite{2019_Amole_PICO}.

\begin{figure*}[htb!]
    \centering
    \includegraphics[width=\textwidth]{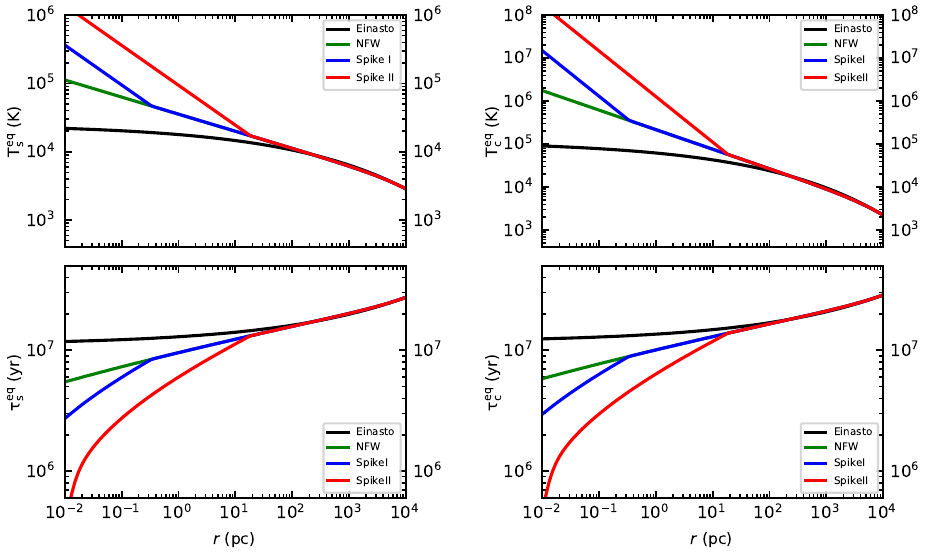}
    \caption{
    Equilibrium thermal properties of NSs located at different galactic radii, as supported by DM heating. The top panels show the equilibrium surface temperature $T_\mathrm{s}^\mathrm{eq}$ (left) and equilibrium core temperature $T_\mathrm{c}^\mathrm{eq}$ (right). The bottom panels present the corresponding times $\tau^\mathrm{eq}$ required to reach thermal equilibrium for the surface (left) and the core (right). 
    Different curves correspond to different DM halo profiles: Einasto~(black), NFW~(green), Spike I~(blue), Spike II~(red).
    }
    \label{fig:relation}
\end{figure*}

\subsection{NS cooling curves} 
With the above physical setup, the initial conditions for the thermal evolution are specified. The initial core temperature is set to $T_{\mathrm{c}}(t=0) = 10^{10}~\mathrm{K}$, and it is assumed that all DM annihilation energy is deposited in the NS. Using these initial conditions, the cooling equation ~\eqref{eq:dT_dt_evo} is numerically solved to obtain the time evolution of $T_{\mathrm{c}}$. 

Figure~\ref{fig:cooling-curve} shows the cooling curves of NSs over a period of $10^8$ years. The colors indicate different NS locations: green corresponds to $0.1~\mathrm{pc}$, blue to $1~\mathrm{pc}$, and red to $10~\mathrm{pc}$. The black line serves as a baseline, representing the standard cooling without DM heating. From left to right, the panels correspond to the Einasto, NFW, Spike~I, and Spike~II DM density profiles. As shown in the top and bottom panels, $T_{\mathrm{c}}$ initially exceeds $T_{\mathrm{s}}$ by approximately two orders of magnitude. As the NS cools, this temperature difference gradually decreases. During the first $\sim 10^3$ years, the NS undergoes rapid cooling dominated by modified Urca neutrino emission. Subsequently, photon emission from the stellar surface becomes the dominant cooling mechanism. Eventually, the NS evolves toward a steady state in which radiative cooling is balanced by heating from DM annihilation. 

The resulting equilibrium temperature $T^{\mathrm{eq}}$ depends sensitively on the ambient DM density. Steeper DM density profiles not only result in higher equilibrium temperatures but also produce a stronger radial dependence of the NS temperature. For the Einasto DM profile, the heating effect is the weakest, and the NS reaches $T_{\mathrm{s}} \sim 10^4~\mathrm{K}$ within the central region considered here. However, in cases of steeper DM density profiles, the heating effect is significantly enhanced. In the most extreme scenario, a NS located at $0.1~\mathrm{pc}$ from the GC with a Spike II DM density profile can reach a $T_\mathrm s$ exceeding $10^5~\mathrm{K}$. By contrast, in the absence of energy injection from DM annihilation, the NS continues to cool without approaching $T^\mathrm{eq}$. 

From the cooling curves in Fig.~\ref{fig:cooling-curve}, we extract $T^{\mathrm{eq}}$ and the corresponding equilibration time ($\tau^{\mathrm{eq}}$), defined as the epoch at which the NS temperature approaches a constant value, indicating a balance between heating and cooling. Figure~\ref{fig:relation} summarizes the equilibrium thermal properties of NSs at various galactic radii for four DM density profiles.

The top panels display the equilibrium surface ($T_{\mathrm{s}}^{\mathrm{eq}}$, left) and core temperatures ($T_{\mathrm{c}}^{\mathrm{eq}}$, right) for NSs located at different galactic radii. At large radii ($r \gtrsim 100~ \mathrm{pc}$), the differences in $T^{\mathrm{eq}}$ and $\tau^{\mathrm{eq}}$ among different DM density profiles become negligible. In contrast, a pronounced divergence emerges in the inner region ($r \lesssim 10~\mathrm{pc}$). A NS located at $r = 0.01~\mathrm{pc}$ and embedded in the Spike~II DM profile reaches a $T_{\mathrm{s}}^{\mathrm{eq}}$ approximately two orders of magnitude higher than that in the Einasto case.
The bottom panels present the corresponding $\tau^{\mathrm{eq}}$ required to reach thermal equilibrium. Depending on the adopted DM density profile, NSs typically reach thermal equilibrium on timescales of $10^6$--$10^7~\mathrm{yr}$. Higher $T^{\mathrm{eq}}$ values correspond to shorter $\tau^{\mathrm{eq}}$. For a Spike~II profile at $r \sim 0.01~\mathrm{pc}$, where the heating is most intense, $\tau^{\mathrm{eq}}$ is as short as $\sim 10^6~\mathrm{yr}$.

\begin{figure*}[htb]
    \centering
    \includegraphics[width=\textwidth]{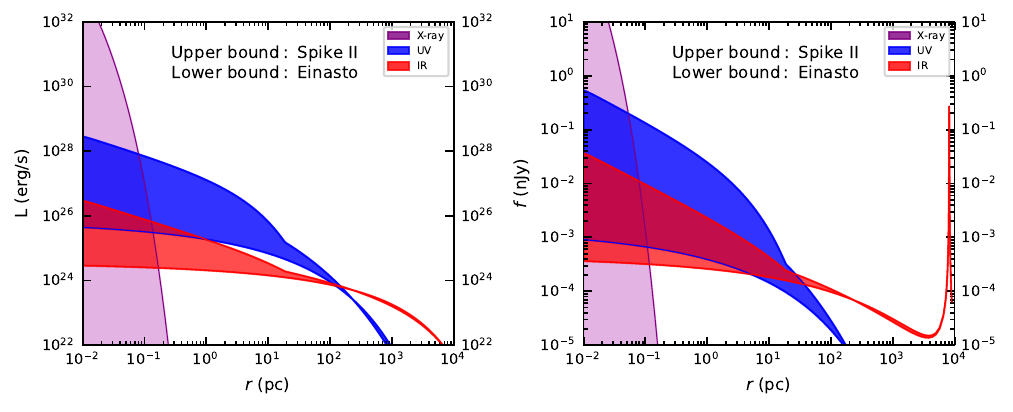}
    \caption{Luminosity $L$ (left panel) and the blackbody radiation flux density $f_{\nu}$ (right panel) as received at Earth from NSs at different galactic radii. Results are shown for different energy bands. The shaded regions represent the predicted ranges of NS luminosity in the x-ray (purple, $0.5$--$2~\mathrm{keV}$), ultraviolet (UV, blue, $0.147$--$0.217~\mu\mathrm{m}$), and infrared (IR, red, $0.624$--$0.781~\mu\mathrm{m}$) bands. For the flux density calculation, representative energies or wavelengths are adopted: the x-ray flux density is evaluated at $0.5~\mathrm{keV}$, and the UV and IR flux densities are evaluated at the center wavelengths of the adopted bands, $0.182~\mu\mathrm{m}$ and $0.705~\mu\mathrm{m}$, respectively. The upper and lower boundaries of each shaded region correspond to the predictions obtained with the Spike II and Einasto DM density profiles, respectively.}
    \label{fig:Landflux}
\end{figure*}

At $t \gtrsim 10^6~ \mathrm{yr}$, neutrino cooling becomes negligible and photon emission dominates. Consequently, the NS evolves toward an equilibrium state in which the photon luminosity approximately balances the heating power supplied by DM annihilation. 
Assuming $\epsilon_\chi \simeq \epsilon_{\gamma}$, the equilibrium surface temperature can be approximated as\begin{equation}
    T_{\mathrm s}^{\mathrm{eq}} \simeq \left(\frac{m_\chi C_{\mathrm{tot}}}{4\pi R^2_{\mathrm{NS}}\sigma_{\mathrm{SB}}}\right)^{1/4}.
\end{equation}
From the definition of $C_{\mathrm{tot}}$ in Eq.~\eqref{eq:single_capture} and using the above relation, the $m_\chi$ dependence cancels out, implying that $T_{\mathrm s}^{\mathrm{eq}}$ is independent of the DM mass. For scattering cross sections below the NS geometric saturation limit, the linear dependence $C_{\mathrm{tot}} \propto \sigma_{\chi_\mathrm{B}}\rho_{\chi}$ implies a scaling of $T_{\mathrm s}^\mathrm{eq} \propto (\sigma_{\chi_\mathrm{B}}\rho_{\chi})^{1/4}$. Here, we adopt $\sigma_{\chi_\mathrm{B}}=10^{-45}~\mathrm{cm^2}$ close to the geometric saturation limit to ensure substantial capture efficiency. Consequently, although the DM density profiles vary by orders of magnitude, the resulting disparity in $T_{\mathrm{s}}^\mathrm{eq}$ is significantly suppressed by this power-law dependence. This analytical estimate is consistent with our numerical results.

\subsection{NS luminosity and flux density}

As mentioned above, an isolated NS remains at a constant temperature when DM heating balances radiative cooling. Assuming the NS emits thermal radiation as a blackbody, the luminosity in a frequency band $[\nu_1,\nu_2]$ is given by
\begin{equation}
L = 4\pi R_{\mathrm{NS}}^2\int_{\nu_1}^{\nu_2} \pi B(\nu,T_{\mathrm{s}})\,d\nu.
\label{eq:Luminosity}
\end{equation}
The Planck specific intensity is defined as \begin{equation}
    B(\nu,T) = \frac{ 2h\nu^3/c^2}{\exp\left(\frac{h\nu}{k_{\mathrm{B}}T}\right)-1}.
\end{equation}
The factor $\pi B(\nu,T)$ corresponds to the radiative flux emitted per unit surface of the NS at frequency $\nu$. To connect with observations, we further consider the spectral flux density received at Earth. Under the blackbody assumption, the flux density is given by
\begin{equation}
    f_{\nu}(r) = \pi B(\nu,T_{\mathrm{s}})\left(\frac{R_{\mathrm{NS} }}{D}\right)^2,
\label{eq:flux_jy}
\end{equation}
where $D$ is the distance between the NS and Earth.

Figure~\ref{fig:Landflux} presents the $L$ (left panel) and $f_\nu$ (right panel) of NSs located at different galactic radii from the GC in three representative observation bands: x-ray (purple), ultraviolet (UV, blue), and infrared (IR, red). The shaded regions indicate the range spanned by the DM profiles considered, with the upper and lower boundaries corresponding to the Spike II and Einasto models, respectively. 

In the innermost region ($r \lesssim 0.1 ~\mathrm{pc}$), efficient DM heating maintains a high $T_{\mathrm{s}}^{\mathrm{eq}}$, leading to emission dominated by soft x-rays. We compute the x-ray luminosity in the $0.5$--$2~\mathrm{keV}$ band and evaluate the flux density at $0.5~\mathrm{keV}$, consistent with the Chandra/ACIS sensitivity. The predicted x-ray luminosity can reach up to $\sim10^{33}~\mathrm{erg\,s^{-1}}$ for the Spike II profile, but declines rapidly with increasing galactic radii.

To assess detectability, we estimate the expected x-ray count rate using the Chandra PIMMS tool (v4.15, Cycle 28)~\footnote{\url{https://cxc.harvard.edu/toolkit/pimms.jsp}}. Modeling the NS emission as a blackbody with $k_{\mathrm{B}}T_{\mathrm{s}} = 0.11~\mathrm{keV}$, we obtain an unabsorbed flux of $6.92\times10^{-14}~\mathrm{erg\,cm^{-2}\,s^{-1}}$ in the $0.5$--$2~\mathrm{keV}$ band. This estimate corresponds to a NS located at $r=0.01~\mathrm{pc}$, assuming the Spike II DM density profile. Adopting a hydrogen column density of $N_\mathrm{H}=8\times10^{22}~\mathrm{cm}^{-2}$ and using the ACIS-I detector over the $0.5$--$8~\mathrm{keV}$ range, the predicted count rate is only $\sim3\times10^{-7}~\mathrm{counts\,s^{-1}}$, several orders of magnitude below the Chandra/ACIS-I point-source sensitivity for any realistic exposure time.  

In the intermediate region ($0.1 \lesssim r \lesssim 10~\mathrm{pc}$), the UV band dominates the emission, with a luminosity roughly an order of magnitude higher than the IR band. This behavior results from the decline of the NS temperature with increasing galactic radii, which shifts the blackbody peak toward longer wavelengths. We adopt the Hubble Space Telescope F25CN182 filter ($0.147$--$0.217~\mu\mathrm{m}$) as a representative UV band. However, strong interstellar extinction toward the GC is expected to severely attenuate the observed UV flux density. 

For NSs located far from the GC, beyond $R_\mathrm{sp}$, DM heating is significantly reduced due to the comparatively low DM density. Consequently, $T_{\mathrm{s}}$ decreases, and the emission becomes dominated by infrared wavelengths. In this regime, we focus on the JWST/NIRCam F070W filter ($0.624$--$0.781~\mu\mathrm{m}$) to characterize the infrared emission. At the central wavelength of the filter ($0.705~\mu\mathrm{m}$), the predicted flux density varies between $\sim10^{-1}$ and $10^{-5}~\mathrm{nJy}$, depending on the DM density profile and galactic radii. Even in the most optimistic cases, the resulting infrared fluxes remain density below the detection limits of current infrared facilities. 

Since the flux scales inversely with the square of the distance [Eq.~(\ref{eq:flux_jy})], the flux density exhibits a peak near $r=8.2~\mathrm{kpc}$ in the right panel of Fig.~\ref{fig:Landflux}. This corresponds to a NS located at a distance of $\sim 15~\mathrm{pc}$ from Earth. At this position, DM heating raises $T_{\mathrm s}$ to $\simeq3000~\mathrm{K}$, producing an infrared flux density of order $10^{-1}~\mathrm{nJy}$ at this wavelength. At longer infrared wavelengths and for nearby neutron stars, the predicted flux density may reach the $\mathrm{nJy}$ level, potentially approaching the sensitivity limits of JWST.

\subsection{Cumulative emission from the NS population} 

Since the flux density from an individual NS is generally too faint to be detected, it is natural to consider the cumulative emission from the NS population in the GC region. However, the number of NSs in the GC remains uncertain. According to general theories of star formation, the GC hosts a dense nuclear star cluster composed of massive stars~\cite{2020_SCH_cluster}, which are expected to evolve into NSs. Based on the observed density of massive stars, Pfahl and Loeb estimated that $\sim 100$--$1000$ pulsars with semimajor axes $\lesssim 0.02~\mathrm{pc}$ should reside in the GC, assuming a pulsar birth rate of $10^{-6}$--$10^{-5}\,\mathrm{yr}^{-1}$~\cite{2004_Pfah1_thousandpulsar}. However, until 2013, only six pulsars have been detected within $\sim 40~\mathrm{pc}$ of Sgr $\mathrm{A}^*$~\cite{2006_Johnston_dictwo, 2009_Deneva_three, 2013_Kennea_magetar, 2025_SHAO_Abbate_SKAO}. Among them, the pulsar closest to the GC, the magnetar PSR J1745-2900, is located within $0.1~\mathrm{pc}$ of Sgr $\mathrm{A}^*$~\cite{2013_Kennea_magetar}. Recently, a reprocessing of the Southern High Time Resolution Universe survey led to the discovery of the pulsar PSR J1746-2829, located within $0.5^\circ$ of Sgr $\mathrm{A}^*$~\cite{2024_Wongphechauxsorn_HTRUS}. Additionally, a millisecond pulsar (MSP) embedded in a radio filament was first detected within $1^\circ$ of Sgr $\mathrm{A}^*$~\cite{2024_Marcus_MilliSGa}. Subsequently, a MSP candidate was reported by the Breakthrough Listen search within the inner $1.4'$ of the GC, exhibiting a high dispersion measure~\cite{2026_Perez_PScandidate}. Nevertheless, detailed analyses of existing deep surveys at 5~\cite{2004_Klein_PulsarEffelsberg,2009_Deneva_three} and $14~\mathrm{GHz}$~\cite{2010_Macquart_high-f} indicate that these surveys should have detected approximately ten ordinary pulsars~\cite{2014_Dexter_Peculiar10}. The lack of such pulsars in the GC is commonly referred to as the ``missing pulsar problem~(MPP).''

One important reason for the nondetection is radio-wave scattering in the ionized interstellar medium (ISM) toward the GC, which smears out pulse profiles on a characteristic timescale $\tau_\mathrm{ISM}\approx 10^3\,\nu^{-4}_{\mathrm{GHz}}\,\mathrm{s}$, where $\nu_{\mathrm{GHz}}$ denotes the observing frequency in GHz~\cite{1997_Cordes_Lazio_SCREEN}. If the pulsar's period is shorter than the pulse broadening timescale $\tau_\mathrm{ISM}$, the periodic signal becomes undetectable. Furthermore, a significant fraction of pulsars may be intrinsically radio quiet or not beaming toward us~\cite{2018_Kim_Davies_NSreview_JKAS}, reducing their detectability. Pulsar searches at higher frequencies are more efficient, as they mitigate the effects of scattering. Despite the growing number of high-frequency pulsar surveys~\cite{2010_Macquart_high-f, 2013_Eatough_nullsurvey, 2021_Eatough_Torne_nullsur, 2023_Torne_nullsur}, no convincing pulsar detections have been made. Meanwhile, the updated model describing free electrons in the Galactic ISM, the so-called NE2005~\cite{2026_Ocker_NE2025}, suggests that the scattering toward the GC is weaker than previously predicted~\cite{2002_Cordes_NE2001,2014_Spitler_broadeningmag}. This makes the MPP even more severe.

One possible explanation for the MPP is that dynamical processes, such as natal kicks~\cite{2014_Dexter_Peculiar10,2022_Boodram_kicks} and gravitational interactions~\cite{2017_Abbate_grivaty}, may eject NSs from the central parsec. However, these mechanisms alone are unlikely to fully account for the MPP, as the expected surviving NS population should still significantly exceed the number of observed pulsars.
An alternative explanation is that the NS population in the GC is dominated by magnetars or recycled MSPs rather than ordinary pulsars~\cite{2014_Dexter_Peculiar10, 2015_Macquart_Pierre_Millisecongd,2024_Bhura_McDonald_Axionpopula}. Magnetars have relatively short active lifetimes and often exhibit intermittent radio activity, while the short spin periods of MSPs make them especially susceptible to temporal smearing by interstellar scattering, thereby reducing their detectability.

Nevertheless, regardless of whether the NS population is dominated by ordinary pulsars, magnetars, or MSPs, DM annihilation-induced heating is expected to affect the thermal evolution of these NSs in a broadly similar manner. DM heating of NSs provides a complementary and potential probe of the NS population in the GC.
\begin{figure*}[htb]
    \centering
      \includegraphics[width=\textwidth]{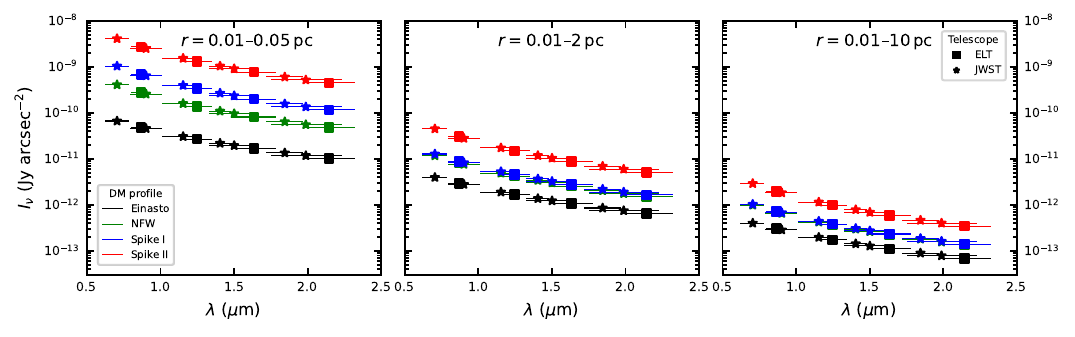}
    \caption{Average surface brightness $I_\nu$ of a population of NSs in the GC, corresponding to the total blackbody radiation flux density. From left to right, the panels show results integrated radially from an inner radius of $r = 0.01~\mathrm{pc}$ out to $0.05~\mathrm{pc}$, $2~\mathrm{pc}$, and $10~\mathrm{pc}$, respectively. Different colors represent various DM density profiles: Einasto~(black), NFW~(green), Spike I~(blue), Spike II~(red). Horizontal bars indicate the bandwidths of the filters, while symbols denote the center wavelengths at which the surface brightness is evaluated, with stars corresponding to JWST and squares to ELT.} \label{Figure:integrate_NS_telescope_combine}
\end{figure*}
The number density profile of NSs was modeled by Generozov et al.~\cite{Generozov_2018} using time-dependent Fokker-Planck (FP) models of nuclear star cluster dynamics, which self-consistently account for two-body relaxation and continuous injection of compact remnants. The fiducial $\times~10$ model \cite{Generozov_2018} assumes a NS formation rate of $\dot N_{\mathrm {NS}}=4\times10^{-5}~\mathrm{yr}^{-1}$, corresponding to the present-day formation rate of massive stars in young stellar disks. Under these assumptions, the resulting spatial number density distribution of NSs is approximated by a broken power-law profile,
 \begin{align}
 n_{\mathrm{NS}}(r)=
\begin{cases}
  5.98\times 10^3\left(\frac{r}{1~\mathrm{pc}}\right)^{-1.7} \mathrm{pc}^{-3}&(0.1~\mathrm{pc}< r <2~\mathrm{pc} ),\\[6pt]
  2.08\times 10^4 \left(\frac{r}{1~\mathrm{pc}}\right)^{-3.5} \mathrm{pc}^{-3}&(r>2~\mathrm{pc}).
\end{cases}
\label{NSdensity}
\end{align}
 For simplicity, we assume the NS population in the GC region to lie at a common distance $D=D_{\mathrm{GC}}=8.2~\mathrm{kpc}$. The total flux density received at Earth from a population of NSs within a spherical shell $r_1 < r <r_2$ is obtained by integrating the individual contributions over the radial distribution,
\begin{equation}
    F_{\mathrm{pop}}(\nu)=4 \pi\int_{r_1}^{r_2}r^2\,n_{\mathrm{NS}}(r) f_\nu(r)\,dr.
\end{equation}
To facilitate comparison with observational sensitivities, the corresponding average surface brightness can be given by
\begin{equation}
    I_{\nu} =\frac {F_{\mathrm{pop}(\nu)}}{\Omega},
\end{equation}
where $\Omega$ is the solid angle subtended by the integration region on the sky. For a spherical shell bounded by radii $r_1$ and $r_2$, the solid angle is approximated as $\Omega \approx \pi(r_2^2-r_1^2)/D_{\mathrm{GC}}^2$.
Figure~\ref{Figure:integrate_NS_telescope_combine} shows the surface brightness $I_\nu$ at the central wavelengths of the adopted JWST/NIRCam (stars) and ELT/MICADO (squares) filters for different DM density profiles. $I_{\nu}$ is obtained by integrating the NS population from an inner radius of $r = 0.01~\mathrm{pc}$ to outer radii of $0.05~\mathrm{pc}$, $2~\mathrm{pc}$, and $10~\mathrm{pc}$, as shown in the left, middle, and right panels, respectively. $I_\nu$ is enhanced for smaller integration areas due to the significant concentration of both DM and NSs in the innermost region. For a fixed integration region within $r < 10~\mathrm{pc}$, $I_\nu$ increases toward shorter wavelengths, consistent with the high $T_{\mathrm{s}}^{\mathrm{eq}}>10^4~\mathrm{K}$ of NSs. As the integration region is extended to large radii, the impact of the central DM spike becomes progressively diluted by the increasing contribution from the outer regions. In particular, for the Spike I profile, the predicted $I_\nu$ converges to the NFW case once the integration radius exceeds the DM spike radius $R_\mathrm{sp}$.

To assess the potential detectability of the NS heating signal, we estimate the expected signal-to-noise ratio (SNR) using the JWST Exposure Time Calculator (ETC, version 5.1)~\footnote{\url{https://jwst.etc.stsci.edu/}}. We adopt an optimistic setup by neglecting interstellar extinction and selecting the shortest-wavelength NIRCam imaging filter, F070W ($0.705~\mu\mathrm{m}$), at which the predicted $I_{\nu}$ is maximal. The NS population is modeled as an extended source, located at the GC coordinates, with $I_\nu=4\times10^{-9}~\mathrm{Jy}\,\mathrm{arcsec}^{-2}$ normalized at $\lambda=0.705~\mu
\mathrm{m}$. Even with an extremely long total exposure time of $\sim 10^6~\mathrm{s}$, the ETC predicts a SNR of only $\simeq 1.14$. Notably, Ref.~\cite{2014_Dexter_Peculiar10} investigates NS natal kicks in the GC, assuming formation in a stellar disk similar to that observed in the central parsec~\cite{2006_Paumard_diskcenpc} and adopting a double-exponential kick velocity distribution~\cite{2006_Faucher_KICKdist}. They find retention fractions of $30\%$ to $60\%$ within $0.1\sim10~\mathrm{pc}$ of the GC. The NS density distribution derived from the FP model used here does not account for natal kicks, and the authors of Ref.~\cite{Generozov_2018} point out that roughly $60\%$ of the initial NS population is expected to remain bound after natal kicks. Since the predicted signal approximately scales with this NS retention fraction, the resulting thermal surface brightness would be correspondingly reduced. For simplicity, natal kicks are not quantitatively included in this work, and thus our results represent an upper limit on the cumulative thermal surface brightness. Galactic extinction would further suppress the observable signal. We therefore conclude that the thermal emission from a population of NSs heated by DM annihilation is unlikely to be detectable with current facilities. 

\section{Conclusion}
In this work, we investigated the heating of NSs induced by the capture and annihilation of DM in the GC. We accounted for both kinetic energy deposition during the capture process and the subsequent annihilation of accumulated DM in the stellar interior. These energy injection mechanisms significantly modify the late-time thermal evolution of NSs. The magnitude of this effect strongly depends on the assumed DM density profile, with a pronounced disparity arising between cored and cuspy distributions near the GC. For NSs older than $\sim10^7~\mathrm{yr}$, $T_{\mathrm{s}}^\mathrm{eq}$ in the inner region is predicted to lie in the range $\sim10^4$ to $10^6~\mathrm{K}$.

We further assessed the observational prospects of this heating scenario by estimating the expected luminosities and flux densities of NSs in the GC. Although the enhanced $T_{\mathrm{s}}$ in the presence of a DM density spike shift the emission toward the UV and soft x-ray bands, interstellar extinction severely suppresses the UV radiation. Even for a soft x-ray luminosity of order $\sim 10^{33}~\mathrm{erg\,s^{-1}}$, the emission is strongly attenuated by the large hydrogen column density toward the GC, yielding count rates far below the detection thresholds of current x-ray facilities. At larger galactic radii, the emission becomes predominantly infrared as the x-ray component rapidly weakens. Even under an extremely optimistic scenario, the predicted IR flux density from an individual NS located near the GC remains below $0.1~\mathrm{nJy}$, rendering direct detection infeasible. Considering the collective emission from a population of NSs, $I_\nu$ reaches at most $10^{-9}~\mathrm{Jy}\,\mathrm{arcsec^{-2}}$ even in the innermost region ($r = 0.01$--$0.05~\mathrm{pc}$). Nevertheless, the corresponding SNR remains well below the detection threshold ($\mathrm{SNR} < 2$) for extremely long exposure times. Combined with the strong astrophysical background in the GC, these results indicate that the thermal emission from DM-heated NS populations is unlikely to be observable with current infrared facilities.   

Although detection in the GC is challenging, the signal could be more favorable in less obscured and closer systems. In particular, nearby dwarf spheroidal galaxies with low hydrogen column densities, as well as proximate globular, may improve the detectability of thermal emission from DM-heated NS populations. Furthermore, the peak flux density shown in Fig.~\ref{fig:Landflux} corresponds to an isolated NS at a distance of $\sim 15~\mathrm{pc}$ from Earth with $T_{\mathrm s}^\mathrm{eq}\sim 3000~\mathrm{K}$. This distance is comparable to the $\sim 20~\mathrm{pc}$ infrared detection horizon estimated in Ref.~\cite{2024_NirmalRaj_ReheatedNS}.
\begin{acknowledgments}
We thank Shu Luo for helpful discussions and Haikun Li for support with the numerical implementation. 
We also thank the anonymous referee for valuable comments that improved the quality of this paper. 
This work is supported by the National SKA Program of China No. 2025SKA0150103, and the National Natural Science Foundation of China under Grants No. 12550002, No. 12133008, No. 12221003, No. 11890692. We acknowledge the science research grants from the China Manned Space Project with Grants No. CMS-CSST-2021-A04 and No. CMS-CSST-2025-A10.
\end{acknowledgments}
\bibliographystyle{apsrev4-2}
\bibliography{reference}

\end{document}